\documentclass{PoS}
\usepackage{amsmath,amssymb,graphicx,enumerate}

\title{O(3) model with Nienhuis action}

\ShortTitle{O(3) model with Nienhuis action}

\author{Ferenc Niedermayer\\
        Albert Einstein Center for Fundamental Physics, Institute for Theoretical Physics, University of Bern, Switzerland\\
        E-mail: \email{niederma@itp.unibe.ch}}

\author{\speaker{Ulli Wolff}%
        \\
       Humboldt Universitat zu Berlin, Institut fur Physik, Newtonstrasse 15, D-12489 Berlin, Germany\\
       E-mail: \email{uwolff@physik.hu-berlin.de}}

\abstract{
We study the O(3) sigma model in $D=2$ on the lattice with a Boltzmann weight linearized in $\beta$ on each link. 
While the spin formulation now suffers from a sign-problem the equivalent loop model remains positive and 
becomes particularly simple. By studying the transfer matrix and by performing Monte Carlo simulations in the loop form 
we study the mass gap coupling in a step scaling analysis. The question addressed is, whether or not such a 
simplified action still has the right universal continuum limit. If the answer is affirmative this would be 
helpful in widening the applicability of worm algorithm methods.
\begin{flushright} HU-EP-16/37 \end{flushright}
}

\FullConference{34th annual International Symposium on Lattice Field Theory\\
                 24-30 July 2016\\
                 University of Southampton, UK}

\newcommand{\Omicron}{\mathrm{O}}
\newcommand{\mathe}{\mathrm{e}}
\newcommand{\tmem}[1]{{\em #1\/}}
\newcommand{\tmop}[1]{\ensuremath{\operatorname{#1}}}
\newcommand{\tmtextbf}[1]{{\bfseries{#1}}}
\newcommand{\tmtextit}[1]{{\itshape{#1}}}
\newcommand{\tmtexttt}[1]{{\ttfamily{#1}}}
\newcommand{\tmtt}[1]{\texttt{#1}}
\newenvironment{enumerateroman}{\begin{enumerate}[i.] }{\end{enumerate}}
\definecolor{grey}{rgb}{0.75,0.75,0.75}
\definecolor{orange}{rgb}{1.0,0.5,0.5}
\definecolor{brown}{rgb}{0.5,0.25,0.0}
\definecolor{pink}{rgb}{1.0,0.5,0.5}

\begin{document}


\section{Introduction}

In recent years the efficient worm algorithm simulation method of Prokof'ev
and Svistunov {\cite{prokofev2001wacci}} has been successfully generalized to
the O($N$) {\cite{Wolff:2009kp}} and CP($\left. N - 1 \right)$
{\cite{Wolff:2010qz}} class of nonlinear sigma models with continuous spin
manifolds. In this approach one samples the complete set of strong coupling
graphs of the model instead of spin configurations. Since this expansion of
the partition function in the inverse coupling $\beta$ is absolutely
convergent for {\tmem{finite lattices}} this is an equivalent representation
of the model. The worm method not only deals with the vacuum graphs of the
partition function itself but also with those of the two point function at an
arbitrary pair of arguments. This enlarged scope of graphs simulated is
essential for the efficiency of the simulation dynamics. In addition it makes
the two point function a natural observable which becomes accessible with very
high precision at the physically interesting large distance. Other observables
may be less easily measurable in the graph representation. Further interest in
the reformulated theory derives from the possibility to simulate with
non-zero chemical potential without running into sign problems
{\cite{Bruckmann:2015sua}}.

In the above papers standard lattice formulations have been employed. They
contain exponential Boltzmann weights on each link and hence arbitrarily high
powers of $\beta$. In the graph formulation this implies that in principle
arbitrarily many lines can pile up on top of each other. A significant
simplification therefore arises for the O($N$) model if we truncate
\begin{equation}
  Z = \int D s\, { \mathe^{\beta \sum_{l = \langle x y \rangle} s (x)
  \cdot s (y)}} \hspace{1em}\rightarrow\hspace{1em}
  Z = \int D s \prod_l {[ 1 + \tilde{\beta} s (x) \cdot s (y)
  ]},
  \hspace{1em} s \in S_{N - 1}. \label{Nie}
\end{equation}
In these formulas $s \left( x \right)$ means an $N$-component spin field on
all sites of a $D$ dimensional lattice and $D s$ implies independent
integrations of the spins over the sphere. Energy bonds arise on links $l$ of
nearest neighbor pairs on a hypercubic lattice in arbitrary dimension. The
truncated model trivially has the same degrees of freedom and
symmetry and may therefore be naively conjectured to lie in the same
universality class (same continuum limit). On the other hand,
for $\tilde{\beta} > 1$, the weight in (\ref{Nie}) oscillates and we leave the
realm of standard statistical physics. In the equivalent graph form no sign
problem appears and we hence use it to numerically investigate universality in
this system. In the affirmative case, the permission to employ such
truncations could be very useful to extend the range of models that can
studied with worm-type algorithms.

The truncated O($N$) model has been intensely studied before by Nienhuis and
collaborators {\cite{Domany:1981fg}}, {\cite{Nienhuis:1982fx}}. They assumed
universality and focused on the range $- 2 \leqslant N \leqslant 2$ on the two
dimensional honeycomb lattice. Here only 3 links surround a site and the graph
structure simplifies even more by not allowing any intersections. As a
consequence the models in the graph representation could be solved exactly,
including the famous XY model $N = 2$. The occurring $\tilde{\beta}$ are
smaller than one and the range above one is called unphysical in these references. 
We add here
that $N = 1$ is the Ising model, and here both actions are
trivially equivalent if we identify $\tilde{\beta} = \tanh \beta$.

\section{Transfer matrix}
Specializing to $D = 2$ in this section we now investigate the truncated transfer matrix
\begin{equation}
  \mathbb{T} =\mathbb{T}_0 \mathbb{T}_1,\hspace{1em}
  \mathbb{T}_0 \left[ s', s \right] = \prod_z \left( 1 + \tilde{\beta} s'
  \left( z \right) \cdot s \left( z \right) \right), \hspace{1em}
  \mathbb{T}_1 \left[ s \right] = \prod_{\langle x y \rangle} \left( 1 +
  \tilde{\beta} s \left( x \right) \cdot s \left( y \right) \right),
  \label{T01}
\end{equation}
where $x, y, z$ here are sites on a row of spins and $s, s'$ denote
configurations thereon. $\mathbb{T}$ acts on states or `wave functions' $\psi
\left[ s \right] \rightarrow \left( \mathbb{T} \psi \right) \left[ s \right]
= \int D s' \mathbb{T} \left[ s, s' \right] \psi \left[ s' \right]$ and we
notice that the image is a product of monomials of zeroth and first order in
$s \left( x \right)$. Hence $\mathbb{T}$ projects to a Hilbert space of
finite dimension $\left( 1 + N \right)^L$ if there are $L$ sites in the row
which we close periodically. This allowed us to straight forwardly set up
$\mathbb{T}$ as a finite matrix and we could numerically explore its spectrum
up to $L = 14$ on a PC and without any special tricks. Since as it stands
$\mathbb{T}$ is real but not symmetric we first symmetrize it by
the similarity transformation
\begin{equation}
  \mathbb{T} \rightarrow \mathbb{T}_0^{- 1 / 2} \mathbb{T}\mathbb{T}^{1 /
  2}_0, \hspace{1em} \tmop{using} \hspace{1em} \mathbb{T}_0^{\gamma} \left[
  s', s \right] = \prod_z \left( 1 + \tilde{\beta}^{\gamma} N^{1 - \gamma} s'
  \left( z \right) \cdot s \left( z \right) \right) .
\end{equation}
Thus we know that $\mathbb{T}$ has real but not necessarily positive
eigenvalues and the two-step matrix $\mathbb{T}^2$ is real positive.

\begin{figure}[htb]
\begin{center}
  \resizebox{0.5\textwidth}{!}{\includegraphics{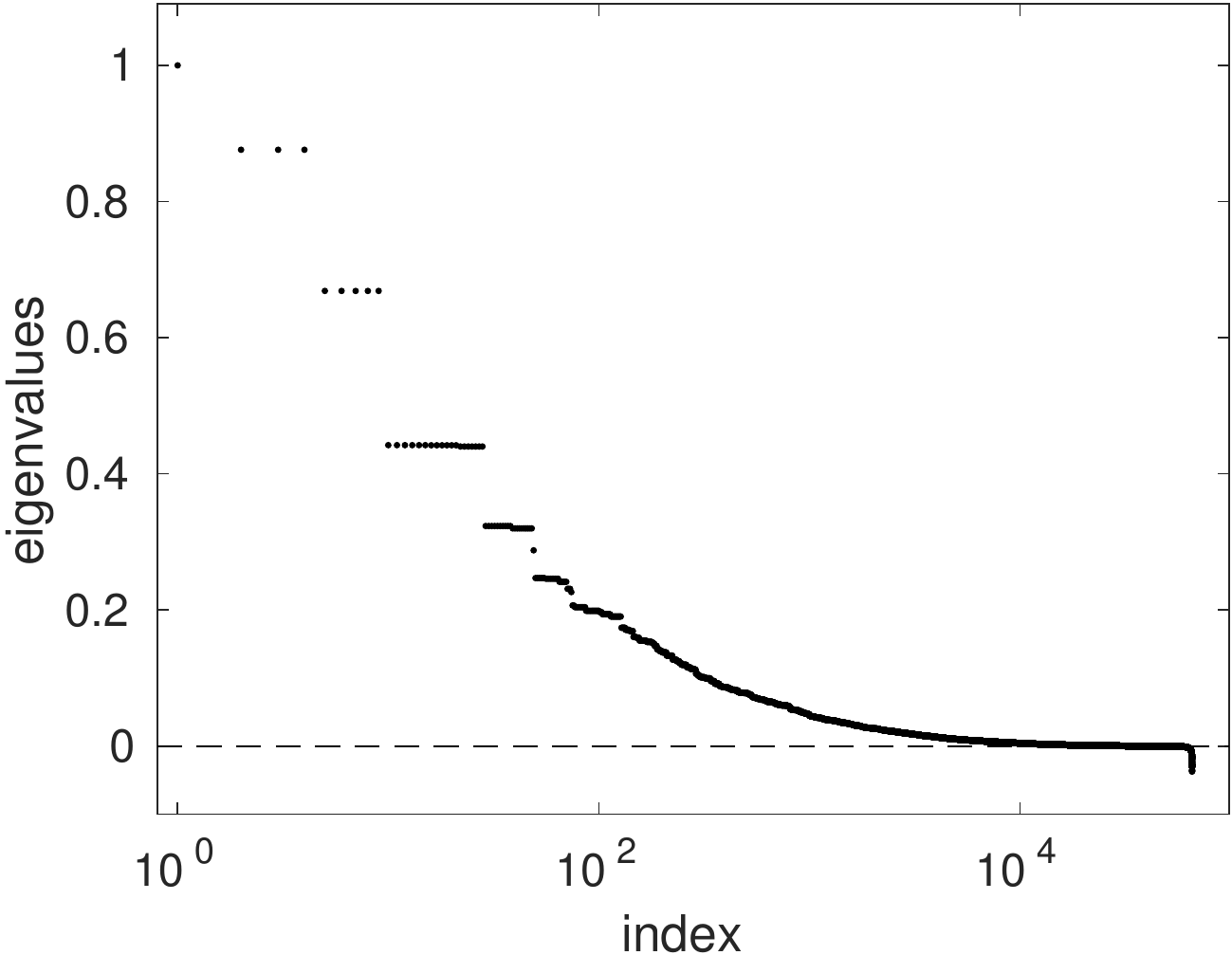}}
  \caption{Complete spectrum of $\mathbb{T}$ in the O(3) model
  for $L = 8$ and $\tilde{\beta} = 1.9144$.\label{full}}
\end{center}
\end{figure}

In Fig. \ref{full} we display a complete spectrum of $\mathbb{T}$ in the O(3)
model for parameters which lead to a finite volume
mass gap $m \left( L \right) L = 1.0595$ which is employed in many
calculations of the step scaling function {\cite{Luscher:1991wu}}. Note, that
this is already in the range where (\ref{Nie}) has a sign problem. The
spectrum was obtained by applying the routine {\tmtt{eig}} under
{\tmtt{matlab}} and it is ordered and plotted logarithmically against the
index. The largest eigenvalue has been normalized to unity and is a single
point, the non-degenerate ground state. As expected the first excitation is a
triplet, presumably at zero momentum. Then many more points follow -- 65536 in
total -- with the last 21404 of them being negative but small in modulus.

\begin{figure}[htb]
\begin{center}
  \resizebox{0.4\textwidth}{!}{\includegraphics{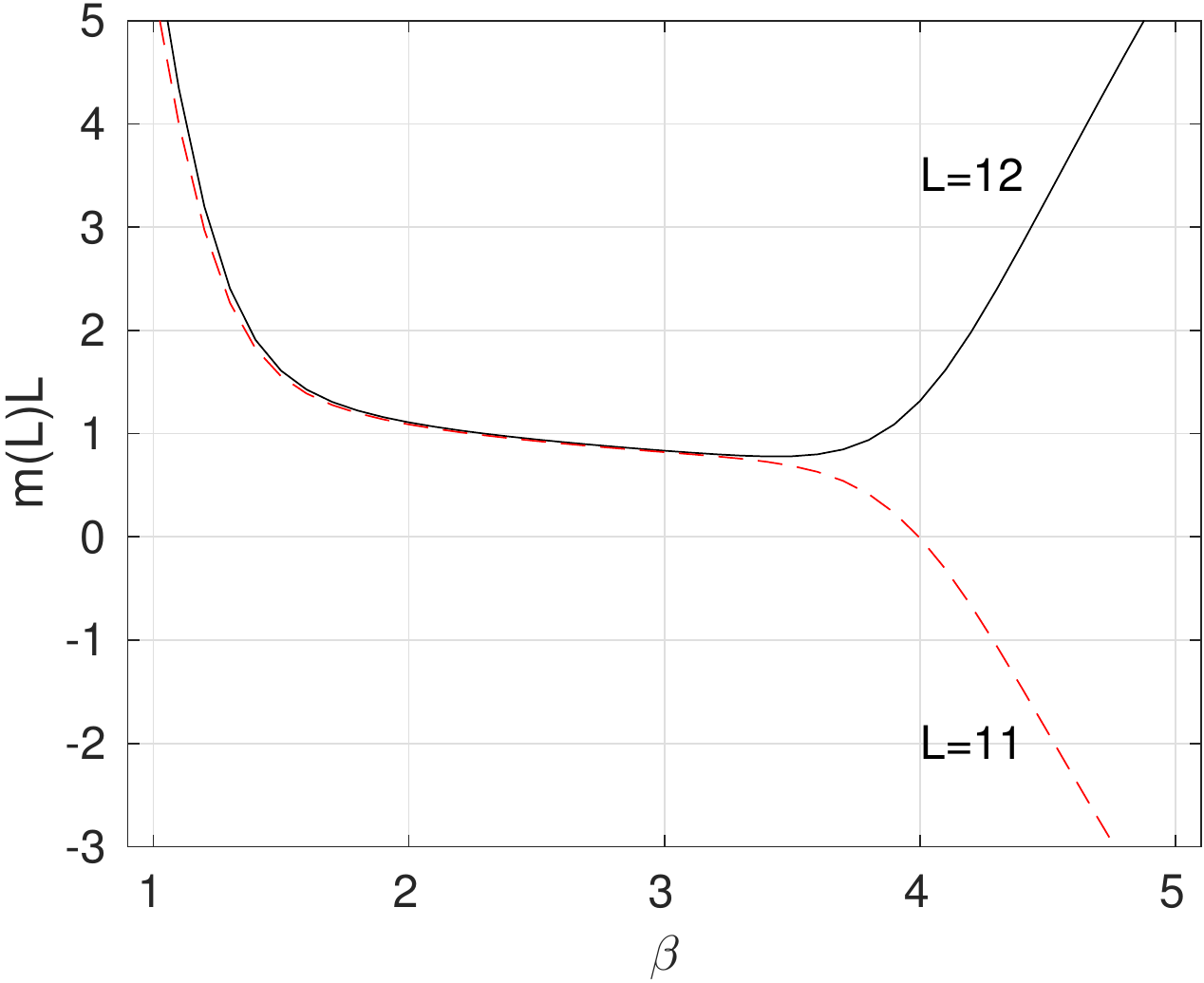}}
\resizebox{0.45\textwidth}{!}{\includegraphics{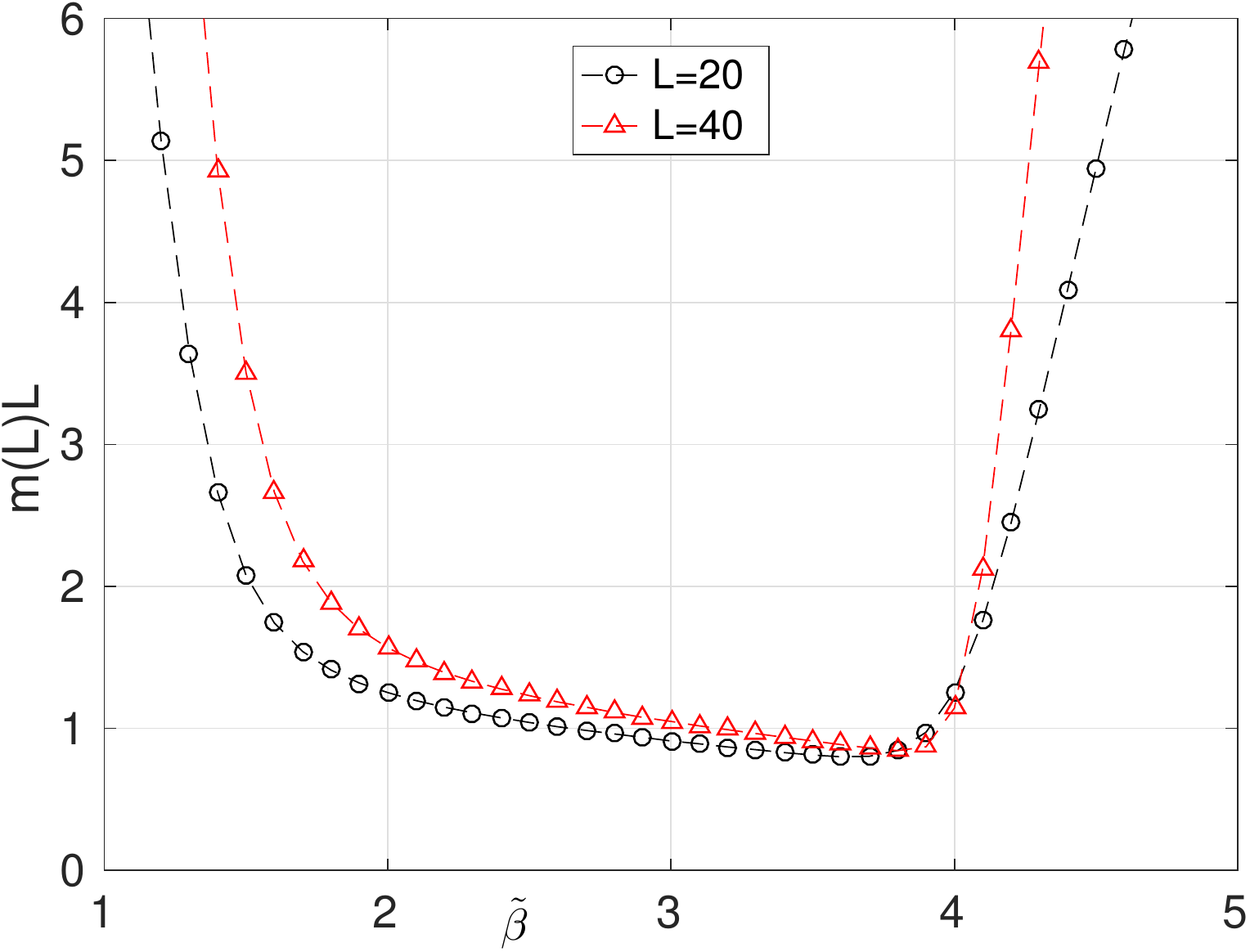}}  
  \caption{Finite volume mass gap against $\tilde{\beta}$. Left panel: typical even
  and odd lattices from the transfer matrix. Right panel: Worm Monte Carlo. \label{mLb}  }
\end{center}  
\end{figure}

In Fig. \ref{mLb} (left) we see the mass gap versus $\tilde{\beta}$ for $L = 12$
and $L = 11$. \ As explained in {\cite{Luscher:1991wu}} the mass gap may be
considered as a scale dependent renormalized coupling constant $m(L)L = \bar{g}^2(L)$.
Concentrating on the even case first, we notice that $\bar{g}^2$ comes
from large values around (and below) $\tilde{\beta} = 1$, reaches a minimum
between 3 and 4 and then rises again. In other words the renormalized coupling
as a function of the bare coupling $\beta^{- 1}$ never gets very small. This
is in contrast to the standard action where the perturbative behavior
$\bar{g}^2 = \beta^{- 1} + c \left( L \right) \beta^{- 2} + \Omicron
\left( \beta^{- 3} \right)$ holds for larger $\beta$.

On $L = 11$ and other odd size lattices one observes a crossing of eigenvalues
of the singlet ground state with the triplet, and as we have set $\exp \left(
- m \right) = \lambda_1 / \lambda_0$ the mass changes sign where the triplet
becomes the states with the largest eigenvalue of $\mathbb{T}$. This symmetry
breaking corresponds to a violation of the Mermin Wagner theorem which
apparently does not apply with the action (\ref{Nie}). The dominance of a non
singlet at large $\tilde{\beta}$ can be qualitatively understood: at large
$\tilde{\beta}$ it is favorable to always insert $\tilde{\beta} s' \cdot s$ in
(\ref{T01}) and not the unity term. In the limit a number of $L$ spins have to
be combined which precludes the formation of an over-all singlet in the case
of odd $L$. This clearly is an artifact of the locally (per site) finite
dimensional Hilbert space caused by our truncation.

\section{Worm simulation}

For each truncated bond on link $l = \langle x y \rangle$ we insert the
trivial identity
\begin{equation}
  \left[ 1 + \tilde{\beta} s (x) \cdot s (y) \right] = \sum_{k \left( l
  \right) = 0}^N \delta_{k \left( l \right), 0} + \left( 1 - \delta_{k \left(
  l \right), 0} \right) \tilde{\beta} s_{k \left( l \right)} (x) s_{k \left( l
  \right)} (y)
\end{equation}
which leads to a new field $k \left( l \right) \in \left\{ 0, 1, \ldots, N
\right\}$. For a given configuration $k$ the spin components appear in a giant
monomial and can be integrated out. In this way we can write
\begin{equation}
  \mathcal{Z}= \sum_{u, v, c} \rho^{- 1} \left( u, v \right) \int D s \prod_l
  { \left[ 1 + \tilde{\beta} s (x) \cdot s (y) \right]} s_c \left(
  u \right) s_c \left( v \right)
\end{equation}
as the exactly equivalent graph or loop gas ensemble
\begin{equation}
  \mathcal{Z}= \sum_{u, v, c, \left\{ k \right\}} \rho^{- 1} \left( u, v
  \right)  \tilde{\beta}^{\sum_l \left( 1 - \delta_{k \left( l \right), 0}
  \right)}  \prod_z C \left[ q \left( z \right) \right] . \label{lgas}
\end{equation}
A configuration is visualized by drawing lines of $N$ different colors on
those links that carry one of the $N$ values $k \left( l \right) > 0$. At
sites $z$
\begin{equation}
  q_{{ a}} \left( { z} \right) = \sum_{l, \partial l
  { \ni z}} \delta_{k(l),a} + \delta_{{ a}, c}
  \left( \delta_{{ z}, u} + \delta_{{ z}, v} \right)
  \hspace{2em} \left( a = 1, 2, \ldots, N \right)
\end{equation}
counts the power of the spin component $s_a \left( z \right)$ appearing in the
monomial at site $z$, and upon integration over the sphere (see e.g.
{\cite{Wolff:2009kp}}) we get ($n!! = 0$ for even $n$)
\begin{equation}
  C \left[ q \right] = \int d \mu \left( s \right)  \prod^N_{a = 1} \left( s_a
  \right)^{q_a} = \frac{\prod_a \left( q_a - 1 \right) !!}{N \left( N + 2
  \right) \cdots \left( N + \left| q \left| - 2 \right) \right. \right.},
  \hspace{2em} \left| q \right| = \sum^N_{a = 1} q_a . \label{weiC}
\end{equation}
It is now easy to establish the identity ($V =$ number of sites)
\begin{equation}
  \langle s \left( x \right) \cdot s \left( y \right) \rangle = \rho \left( x,
  y \right) \frac{\langle \delta_{u, x} \delta_{v, y} \rangle}{\langle
  \delta_{u, v} \rangle / V} . \label{2pnt}
\end{equation}
where the expectation value on the right hand side is taken in (\ref{lgas}).
The real positive weight function $\rho$ cancels out but is useful to optimize
the signal to noise ratio {\cite{Wolff:2009kp}}.

To simulate the ensemble (\ref{lgas}) we perform iterations constructed from
the following blend of standard worm steps:
\begin{enumerateroman}
  \item A move of $u$ to one of its nearest neighbors is proposed, which
  defines a link $l$;
  the proposal is immediately rejected unless $k \left( l \right) = 0$ or $k \left( l
  \right) = c$; in this case the proposal includes swapping $k \left( l
  \right)$ between these two values and the proposal is metropolis accepted
  according to the ratio of (finite) weights in (\ref{lgas}).
  
  \item If after a move i. the situation $u = v$ is encountered, a joint move
  to another randomly chosen site is proposed together with a random change of
  $c$ to a new value. Again an accept test follows.
  
  \item An iteration consists of $V$ steps of i./ii. followed by a change of
  the `active' color $c \rightarrow c' \neq c$; this requires us to construct
  a path from $u$ to $v$ travelling through links with $k \left( l \right) =
  c$ which upon traversal are immediately changed to $c'$; if at a site there
  are several possibilities to continue the path, we choose randomly between
  them with equal probabilities. One can prove that the walk necessarily
  arrives at $v$ where it completes. In addition these moves preserve the
  weight and no conditional accept is required. The key to the proof is the observation
  that $\left( q_a - 1 \right) !!$ is precisely the number of ways to pair
  objects of color $a$ at a site and that $\left| q \left| \right. \right.$
  does not change under re-coloring.
\end{enumerateroman}
Steps of type i. may also be applied to $v$, but this is not required. The
combination of i. and ii. alone would be ergodic. However steps of type iii.,
which were discovered by Holger Stephan {\cite{StephanH}}, are essential to
avoid critical slowing down.

\section{Numerical results}

\begin{figure}[htb]
\begin{center}
  \resizebox{0.45\textwidth}{!}{\includegraphics{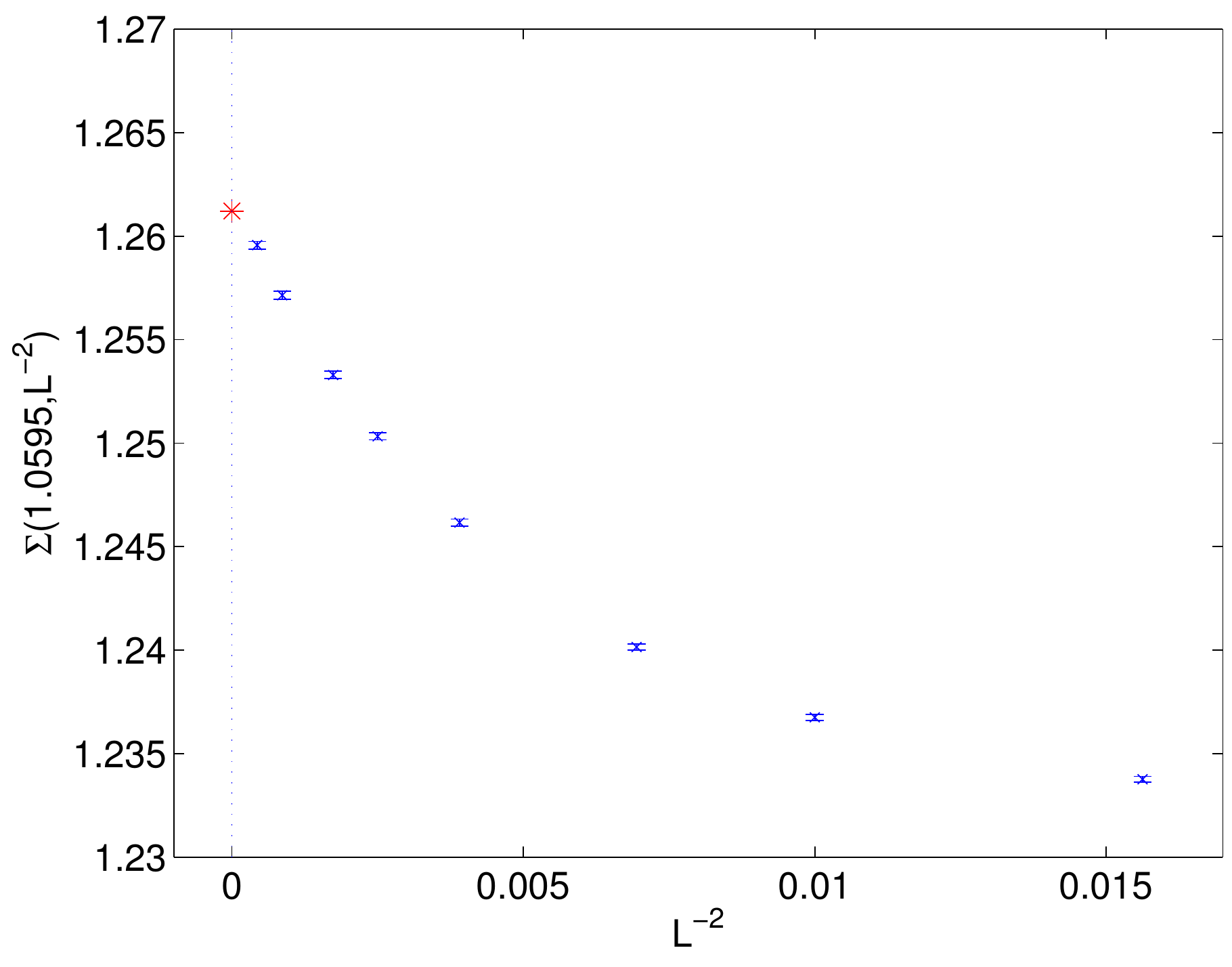}}\resizebox{0.45\textwidth}{!}{\includegraphics{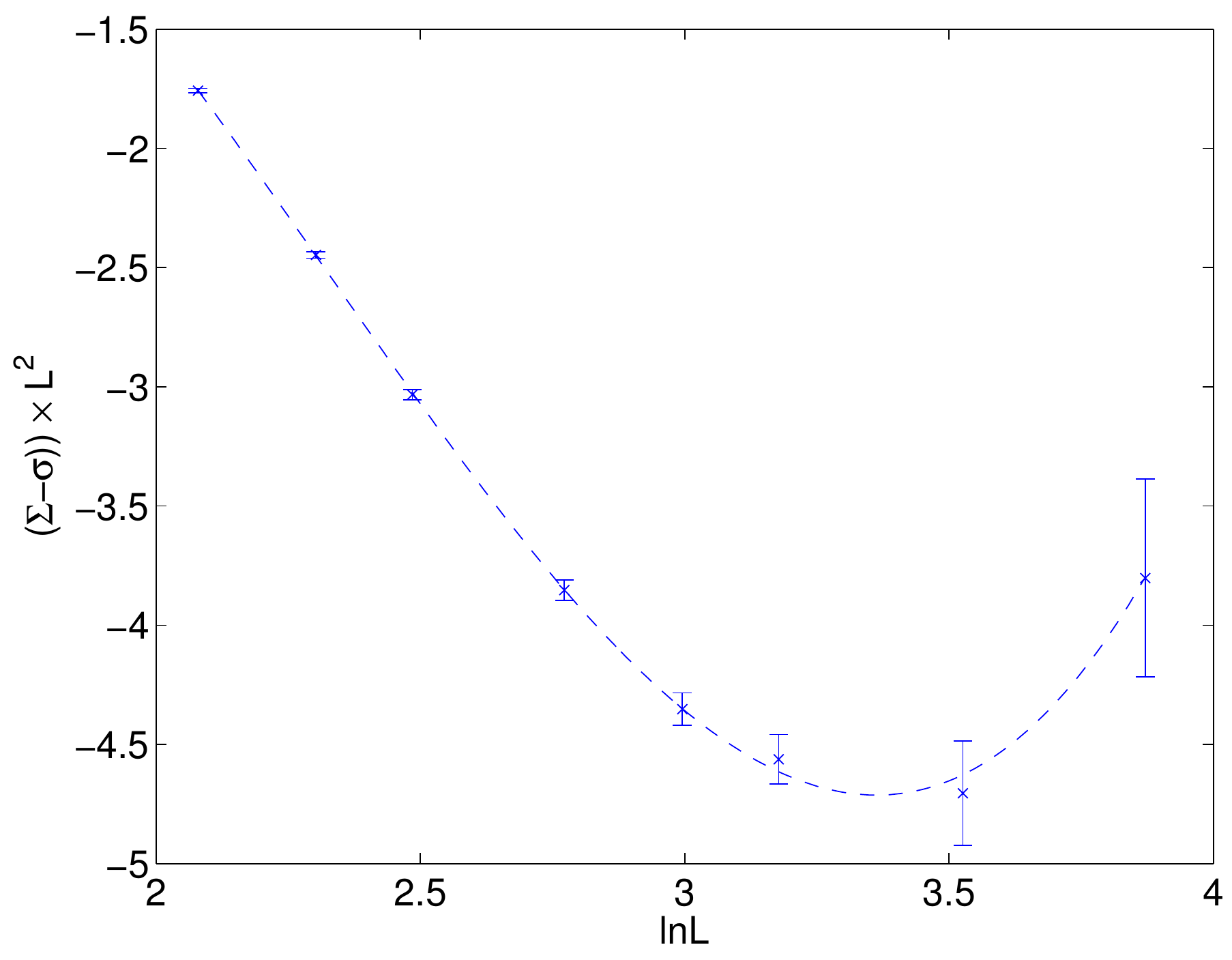}}
  \caption{Extrapolation of $\Sigma \left( 1.0595, L^{- 2} \right)$ to the
  continuum limit with the exact value shown by the red star. Left panel:
  data, right panel: fit of the deviation at finite lattice spacing.
  \label{Sigu0}}
\end{center}   
\end{figure}

Based on (\ref{2pnt}) we determine the time-slice (zero momentum) correlation
and extract the finite volume mass gap by standard techniques
{\cite{Wolff:2009kp}}, {\cite{Luscher:1991wu}}.
In Fig. \ref{mLb} (right) we thus confirm for larger lattices the behavior found
with the transfer matrix. To check universal behavior a convenient tool is the
step scaling function
\begin{equation}
  \sigma(u)= \lim_{L \rightarrow \infty} \Sigma( u, L^{-2}),\quad
  \Sigma(u, L^{- 2}) = \bar{g}^2(2L)_{\bar{g}^2(L) = u}
\end{equation}
where $\sigma$ is known exactly {\cite{Balog:2003yr}} in the asymptotically
free O($N \geqslant 3$) models. In this case $\sigma(u)-u>0$
is the characteristic sign which is true for $\Sigma$ extracted from the pair
of lattices in Fig. \ref{mLb} (right) only in the left part of the figure. We thus
are motivated to tune a number of even sized lattices with $L = 8, \ldots, 48$
to the `traditional' value $\bar{g}^2 \left( L \right) = 1.0595$, which
occur at $\tilde{\beta} = 1.9128, \ldots, 3.0800$, and to then measure
$\bar{g}^2 \left( 2 L \right)$ on the doubled lattices. The result is Fig.
\ref{Sigu0}.
We notice that the data points obviously tend toward the continuum value with
great precision. The fit in the right panel is given by
\begin{equation}
  \Sigma - \sigma = \frac{1}{L^2} \left[ 0.97 \ln^3 L - 6.8 \ln^2 L + 13 \ln L
  - 7.3 \right]
\end{equation}
which is the form derived in {\cite{Balog:2009np}} for a large class of
discretizations which however does not include the Nienhuis action employed
here. Nevertheless it can accommodate our data with competing powers of
logarithms as it has been observed before.

\begin{figure}[htb]
\begin{center}
  \resizebox{0.45\textwidth}{!}{\includegraphics{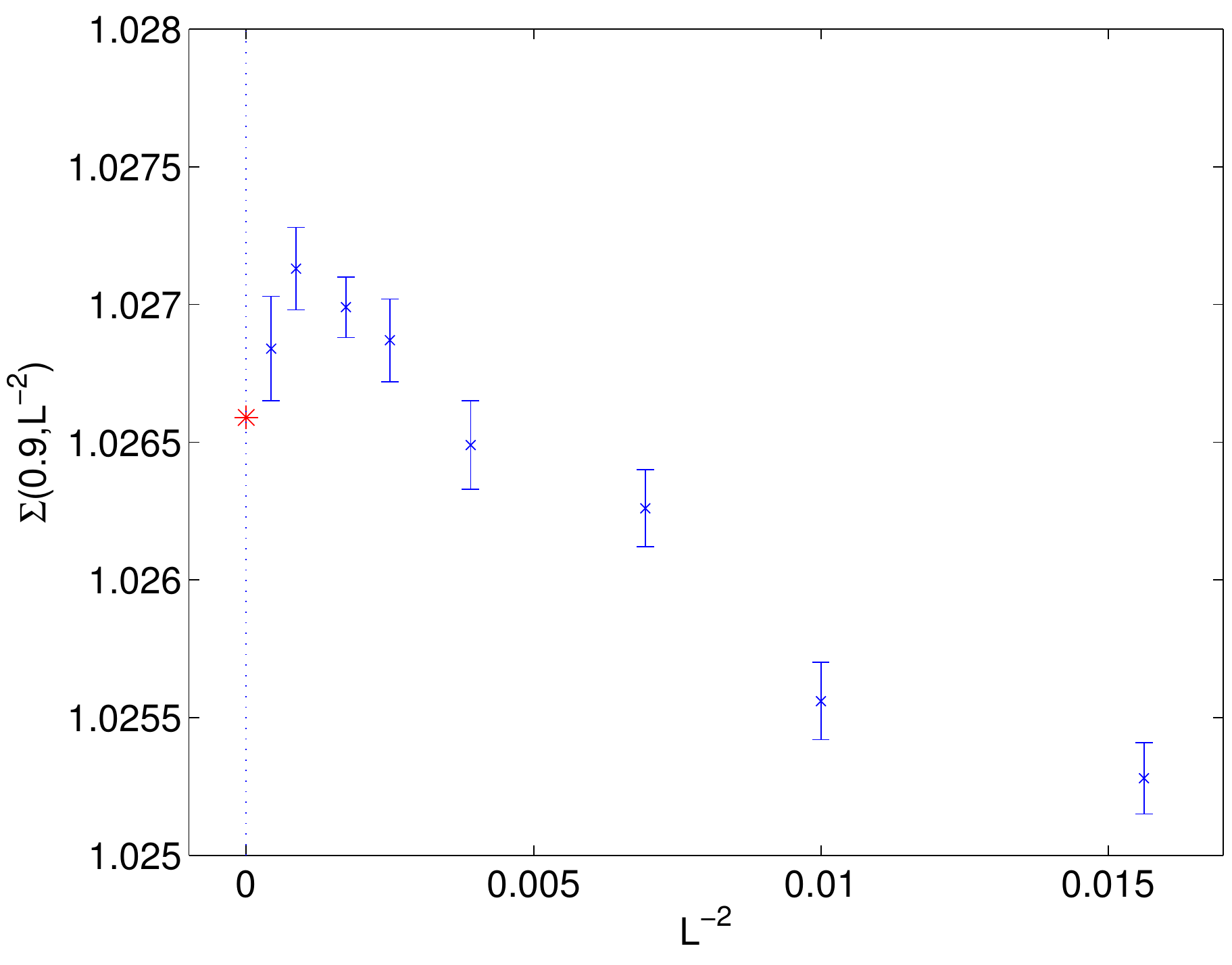}}\resizebox{0.45\textwidth}{!}{\includegraphics{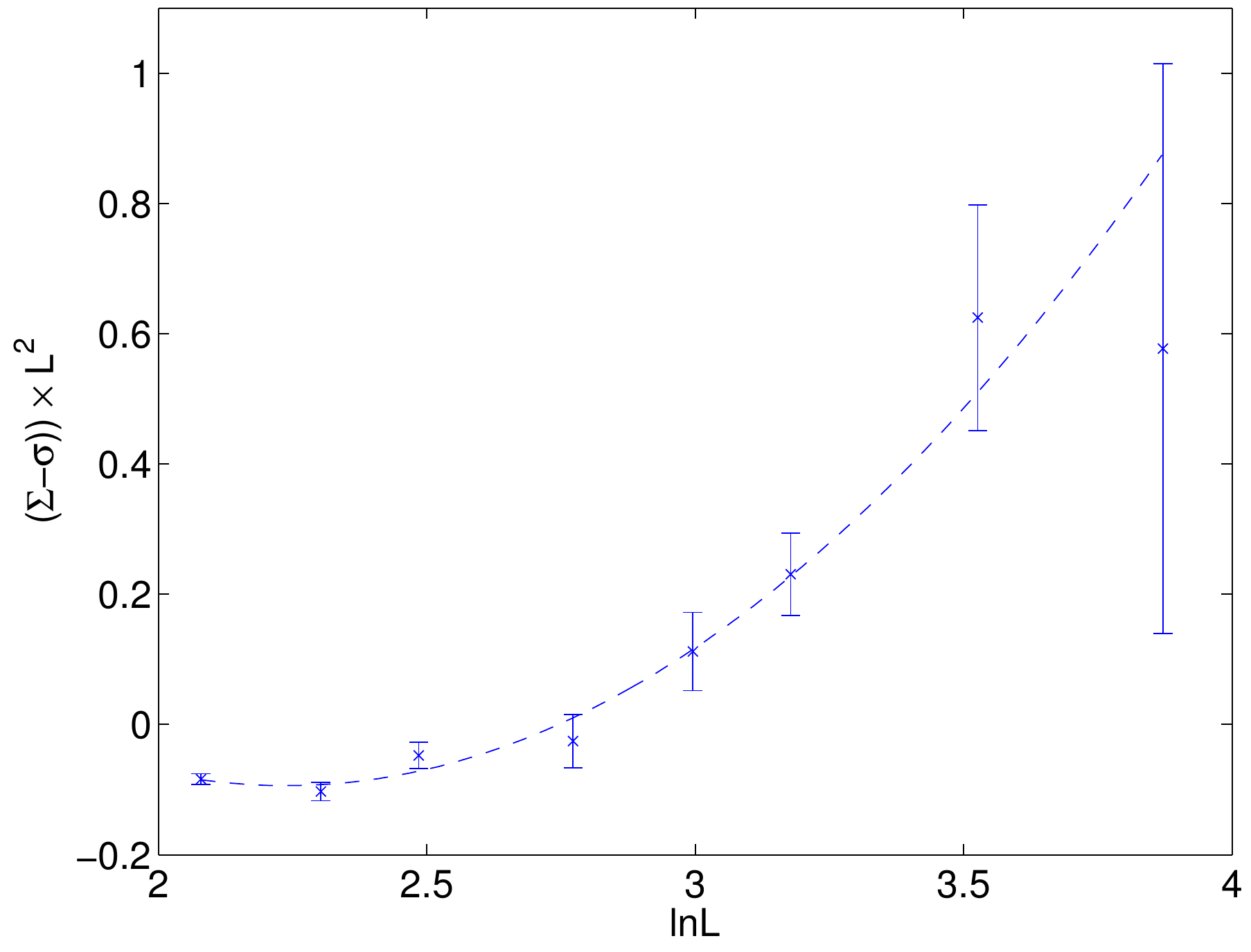}}
  \caption{The same as Fig. \protect\ref{Sigu0} but for $u = 0.9$.\label{Sig0.9}}
\end{center}   
\end{figure}

As witnessed by Fig. \ref{mLb} the range of values $\bar{g}^2$ that can
be realized is bounded toward small couplings. We thus decided to repeat an
extrapolation for $u = 0.9$ which is about the smallest value where reasonably
large $L$ are still possible. This is shown in Fig. \ref{Sig0.9} with fit
coefficients 0, 0.36, $- 1.6$, 1.7 . While the bound obstructs
the use of much larger $L/a$, we notice that on the other hand the lattice
artifacts for the available lattices are particularly tiny here.
As a final remark we mention that in a brief test we verified that
odd size $L$ are of no help to obtain asymptotically free physics for smaller
$mL$. Small values are available at $L$, but $2L$ is even again, and even if we
the average $2L+1$ and $2L-1$ results for the step scaling function are far
away from the universal values.

\section{Conclusions}

We have found in our investigation that the answer to the question
whether the Nienhuis discretization of the 2D O(3) model reproduces
the known universal physics is not a simple yes or no.
Using the well known finite size renormalized coupling based on the mass gap,
we found that for each $\bar{g}^2$ there is a limit $L/a \le f(\bar{g}^2)$
on the available lattice resolutions. In other words, the continuum limit
can not be taken `all the way'. On the other hand, switching to an effective
field theory point of view familiar from many (all?) areas of phyics,
we have demonstrated a very precise approximation of the continuum limit by
rather unambiguosly extrapolating the realizable lattices. As the function $f$
seems to grow with $\bar{g}^2$, the `difficult' end is the perturbative
regime.

\vspace{2ex}

\noindent Acknowledgements: We thank Erhard Seiler and Peter Weisz
for discussions.

\end{document}